\documentclass[twocolumn,aps,prb,superscriptaddress,floatfix,noeprint,longbibliography]{revtex4-1}

\usepackage{graphicx,tabularx}
\usepackage{amsmath,amsfonts,amssymb}
\usepackage{mathtools}
\usepackage{bm,dsfont}
\usepackage[colorlinks=true, citecolor=blue, linkcolor=red, urlcolor=blue]{hyperref}
\usepackage[all]{hypcap}
\usepackage{xcolor}
\usepackage[caption=false]{subfig}
\usepackage[T5]{fontenc}

\AtBeginDocument{%
    \newwrite\bibnotes
    \def\bibnotesext{Notes.bib}
    \immediate\openout\bibnotes=\jobname\bibnotesext
    \immediate\write\bibnotes{@CONTROL{REVTEX41Control}}
    \immediate\write\bibnotes{@CONTROL{%
    apsrev41Control,author="08",editor="1",pages="1",title="0",year="1"}}
     \if@filesw
     \immediate\write\@auxout{\string\citation{apsrev41Control}}%
    \fi
}%

\graphicspath{{figures/}}

\DeclareMathOperator{\sgn}{sgn}

\begin{document}

\title{Network model for periodically strained graphene}
\author{Christophe De Beule}
\affiliation{Department of Physics and Materials Science, University of Luxembourg, L-1511 Luxembourg, Luxembourg}
\affiliation{Department of Physics and Astronomy, University of Pennsylvania, Philadelphia PA 19104}
\author{\fontencoding{T5}\selectfont V\~o Ti\'\ecircumflex{}n Phong}
\affiliation{Department of Physics and Astronomy, University of Pennsylvania, Philadelphia PA 19104}
\author{E. J. Mele}
\affiliation{Department of Physics and Astronomy, University of Pennsylvania, Philadelphia PA 19104}
\date{\today}

\begin{abstract}
The long-wavelength physics of monolayer graphene in the presence of periodic strain fields has a natural chiral scattering network description. When the strain field varies slowly compared to the graphene lattice and the effective magnetic length of the induced valley pseudomagnetic field, the low-energy physics can be understood in terms of valley-polarized percolating domain-wall modes. Inspired by a recent experiment, we consider a strain field with threefold rotation and mirror symmetries but without twofold rotation symmetry, resulting in a system with the connectivity of the oriented kagome network. Scattering processes in this network are captured by a symmetry-constrained phenomenological $S$ matrix. We analyze the phase diagram of the kagome network, and show that the bulk physics of the strained graphene can be qualitatively captured by the network when we account for a percolation transition at charge neutrality. We also discuss the limitations of this approach to properly account for boundary physics.
\end{abstract}

\maketitle

\section{Introduction} \label{sec:intro}

Superlattice engineering in low-dimensional materials by periodic modulations that vary slowly compared to the microscopic lattice has been a highly successful tool for creating artificial crystals that host new emergent phenomena. One route is to impose periodic potentials by patterning electrostatic gates \cite{Forsythe2018} or inducing periodic strain fields with a substrate \cite{Mao2020} or nano-pillars \cite{Jiang2017}. Another platform are moir\'e materials \cite{Andrei2021}, which are layered materials subjected to rotational mismatch or heterostrain, inducing a spatial modulation of the interlayer coupling. The archetypal example is twisted bilayer graphene \cite{LopesDosSantos2007} where the moir\'e structure induces isolated nearly-flat bands at the magic angle that give rise to a rich phenomenology of correlated physics \cite{Cao2018a,Cao2018b}. Similar spectrally isolated and flattened minibands can also emerge in monolayer graphene subjected to a periodic strain field \cite{Mao2020,Milovanovic2020,Manesco2021a,Manesco2021b,Giambastiani2022,Phong2022}.
\begin{figure}
    \centering
    \includegraphics[width=\linewidth]{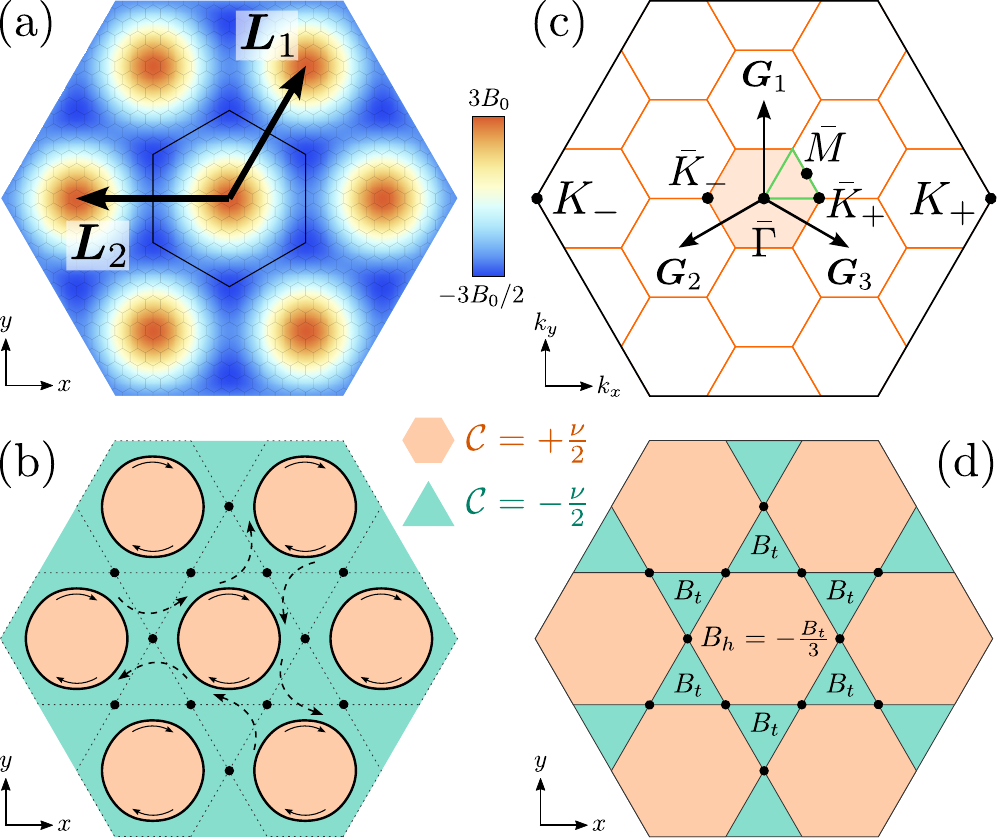}
    \caption{(a) Pseudomagnetic field [Eq.\ \eqref{eq:pmf} for $\phi=0$] superimposed on the graphene lattice for $N=10$. (b) Nodal structure of (a). The PMF is positive (negative) in the orange (green) regions and vanishes on the thick black loops. Each loop supports a chiral mode which can scatter to neighboring loops. Some processes are indicated by the dashed arrows. Here, dotted hexagons highlight the connectivity of the nodal structure with the corners acting as scattering centers (dots). (c) Reciprocal lattice for $N=4$, where the large black hexagon is the original BZ of graphene, and the small orange hexagons give the SBZ in the extended zone scheme. (d) Effective piecewise constant PMF (kagome tiling). The local valley Chern number is indicated for electron doping.}
    \label{fig:system}
\end{figure}

In graphene, a strain field couples dominantly as a vector potential to the long-wavelength Dirac excitations with an opposite sign for the two valleys  \cite{Suzuura2005,Manes2007,Vozmediano2010,Guinea2010,Levy2010}. Hence, a periodic strain field in graphene results in a periodic pseudomagnetic field (PMF), as shown in Fig.\ \ref{fig:system}(a). Wherever the PMF changes sign, which must occur for zero spatially-averaged flux, there are real-space valley Hall transitions. For example, the PMF shown in Fig.\ \ref{fig:system}(a) vanishes on disjoint closed loops as shown in Fig.\ \ref{fig:system}(b). Furthermore, for a PMF that varies slowly on the scale of the graphene lattice and the effective magnetic length, it is meaningful to consider the local Landau levels of regions separated by nodal lines. If the chemical potential lies between the $m$th and $(m+1)$th local Landau level, the net local valley Chern number is given by $\pm (m+1/2)$ since the lowest Landau level of graphene only contributes Chern number 1/2 per valley \cite{Novoselov2005}. Therefore, the local valley Chern number changes by an integer amount when the PMF changes sign. It immediately follows that the nodal lines of the PMF correspond to domain walls that host chiral modes. Due to time-reversal symmetry, these chiral modes propagate in opposite directions for different valleys. However, intervalley backscattering is suppressed when the PMF varies slowly on the scale of the graphene lattice. Furthermore, the bound chiral modes can become dispersive by tunneling to neighboring loops. In our example, the corresponding scattering centers form an effective kagome lattice, as shown in Fig.\ \ref{fig:system}(b). The single-valley low-energy physics of periodically strained graphene can thus be understood in terms of percolating chiral modes with the connectivity of an oriented scattering network. Moreover, for the long-wavelength physics, the smooth PMF can be replaced by an effective piecewise constant field [see Fig.\ \ref{fig:system}(d)] with the same symmetry and connectivity as the smooth PMF.

The above observations illustrate a more general paradigm in which the superlattice modulates the local band topology to give to rise to a Chern mosaic that hosts a chiral scattering network. Similar topological scattering networks occur in moir\'e materials such as minimally twisted bilayer graphene subjected to a perpendicular electric field \cite{San-Jose2013,Efimkin2018,Fleischmann2020,Tsim2020,Chou2020,DeBeule2020a,DeBeule2021,Vakhtel2022} (triangular chiral network), as well as certain large twist-angle graphene bilayers \cite{Pal2019} (helical honeycomb network), and double-aligned graphene-hexagonal boron nitride moir\'es \cite{Moulsdale2022} (chiral kagome network).

In this work, we analyze the scattering network appropriate for periodically strained graphene with zero total flux per valley. We find that the network model captures many attributes of the spectral structure obtained from the valley-projected Hamiltonian. 

This paper is organized as follows. In Section \ref{sec:cont}, we introduce our model for the periodic pseudomagnetic field in graphene and calculate the resulting band structure with a valley-projected continuum theory. We then argue that the low-energy physics can be understood in terms of an effective chiral kagome network. The phenomenological network model is developed in Section \ref{sec:network} where we discuss its phase diagram and corroborate our results by finite-width ribbon calculations. Finally, in Section \ref{sec:disc}, we show that the network model can qualitatively reproduce the bulk low-energy physics of graphene subjected to a periodic strain field and highlight its limitations.

\section{Periodically strained graphene} \label{sec:cont}

To investigate monolayer graphene subjected to periodic strain, we consider a pseudomagnetic field (PMF) with zero net flux that is commensurate with the graphene lattice. Specifically, we consider the case with threefold rotation symmetry ($\mathcal C_{3z}$) about the origin [see Fig.\ \ref{fig:system}(a)] and mirror symmetry across the $y$ axis ($\mathcal M_x$), but with broken twofold rotation symmetry ($\mathcal C_{2z}$). Together with translations, these symmetries form the wallpaper group 14 ($p3m1$) with point group $C_{3v}$. The strain field thus reduces the graphene point group from $C_{6v}$ to $C_{3v}$, which is the case for graphene on top of NbSe$_2$ \cite{Mao2020}. Alternatively, one can engineer this strain field using an artificial substrate with a spatially-varying periodic height profile \cite{Jiang2017,Phong2022}. In the first-star approximation, a PMF with zero net flux that satisfies these symmetry constraints can be written as
\begin{equation} \label{eq:pmf}
    \bm B_\nu(\bm r) = \bm e_z \nu B_0 \sum_{i=1}^3 \cos \left( \bm G_i \cdot \bm r + \phi \right),
\end{equation}
where $\nu=\pm1$ is the valley index, and we take $B_0 > 0$ for concreteness. This PMF is shown in Fig.\ \ref{fig:system}(a) for $\phi = 0$. Note that Eq.\ \eqref{eq:pmf} conserves time-reversal symmetry ($\mathcal T$) since it has an opposite sign in valleys $K_+$ $(\nu=+1)$ and $K_-$ $(\nu=-1)$. We further note that $\mathcal C_{2z}$ is conserved for $\phi = \pi/2$ and maximally broken for $\phi = 0$. Since we are mainly interested in the latter case, we set $\phi = 0$ in the remainder of this work. Here, $\bm G_i$ ($i=1,2,3$) are three of the shortest superlattice reciprocal vectors related by $\mathcal C_{3z}$ symmetry [see Fig.\ \ref{fig:system}(c)]. We further choose the primitive superlattice lattice vectors
\begin{equation}
    \bm L_1 = L(1/2,\sqrt{3}/2), \qquad \bm L_2 = L (-1,0),
\end{equation}
with $L = N a,$ $N > 1$ a positive integer, and $a$ the microscopic lattice constant of graphene [see Fig.\ \ref{fig:system}(a)]. Because the superlattice has a larger periodicity than the original graphene lattice, the graphene Brillouin zone (BZ) is folded onto a smaller superlattice BZ (SBZ) that fits $N^2$ times into the graphene BZ. This is illustrated in Fig.\ \ref{fig:system}(c). For $N \in 3\mathbb N$, the $K_+$ and $K_-$ points are folded on top of each other and the system is an insulator, otherwise the system is a semimetal \cite{Phong2022}. However, for $N \gg 1$, this distinction becomes moot due to an emergent conservation of valley charge. In this case, the PMF varies slowly on the scale of the graphene lattice such that the valleys are effectively decoupled, and one can use a valley-projected continuum theory.

\subsection{Valley-projected theory}

The valley-projected long-wavelength Hamiltonian in the presence of a pseudovector potential is given by
\begin{equation} \label{eq:ham}
    \hat H_\nu = \hbar v_F \int d^2\bm r \, \hat \psi_\nu^\dag \left[ \left( -i\nabla + \frac{e \bm A_\nu}{\hbar} \right) \cdot \left( \nu \sigma_x , \sigma_y \right) \right] \hat \psi_\nu,
\end{equation}
where $\hat \psi_\nu(\bm r) = ( \hat \psi_{\nu A}(\bm r) , \hat \psi_{\nu B}(\bm r) )^t$ are the field operators for sublattices $A$ and $B$, and $\sigma_{x,y}$ are Pauli matrices that act in sublattice space. In dimensionless units, the only free parameter of this theory is given by the ratio $L/l_0$ where $l_0 = \sqrt{\hbar / eB_0}$ is a measure of the magnetic length of the PMF. For example, at the maxima of the PMF, the local magnetic length is given by $l_0/\sqrt{3}$.

The symmetries of the single-valley theory are generated by $\{\mathcal C_{3z}, \mathcal M_x \mathcal T \}$ yielding the magnetic point group $3m$\textquotesingle. Note that twofold rotation $\mathcal C_{2z}$ is not conserved since it sends $\bm B_\nu(\bm r) \mapsto -\bm B_\nu(\bm r)$. In addition, there is a chiral symmetry due to the absence of terms proportional to $\sigma_0$ or $\sigma_z$ in Eq.\ \eqref{eq:ham}. This is a model-dependent symmetry since such terms are symmetry-allowed, e.g., a strain-induced pseudo-electrostatic potential or a constant sublattice-staggering term. However, in this work, we will not consider these terms since they do not qualitatively change our results, as long as they are small compared to $\hbar v_F / l_0$, i.e., the local Landau level splitting, which we assume throughout this work. Since we consider the case where the net flux of the PMF vanishes, the Hamiltonian \eqref{eq:ham} can be readily diagonalized in momentum space. We refer to App.\ \ref{app:cont} for more details on the continuum model and its symmetries.
\begin{figure}
    \centering
    \includegraphics[width=\linewidth]{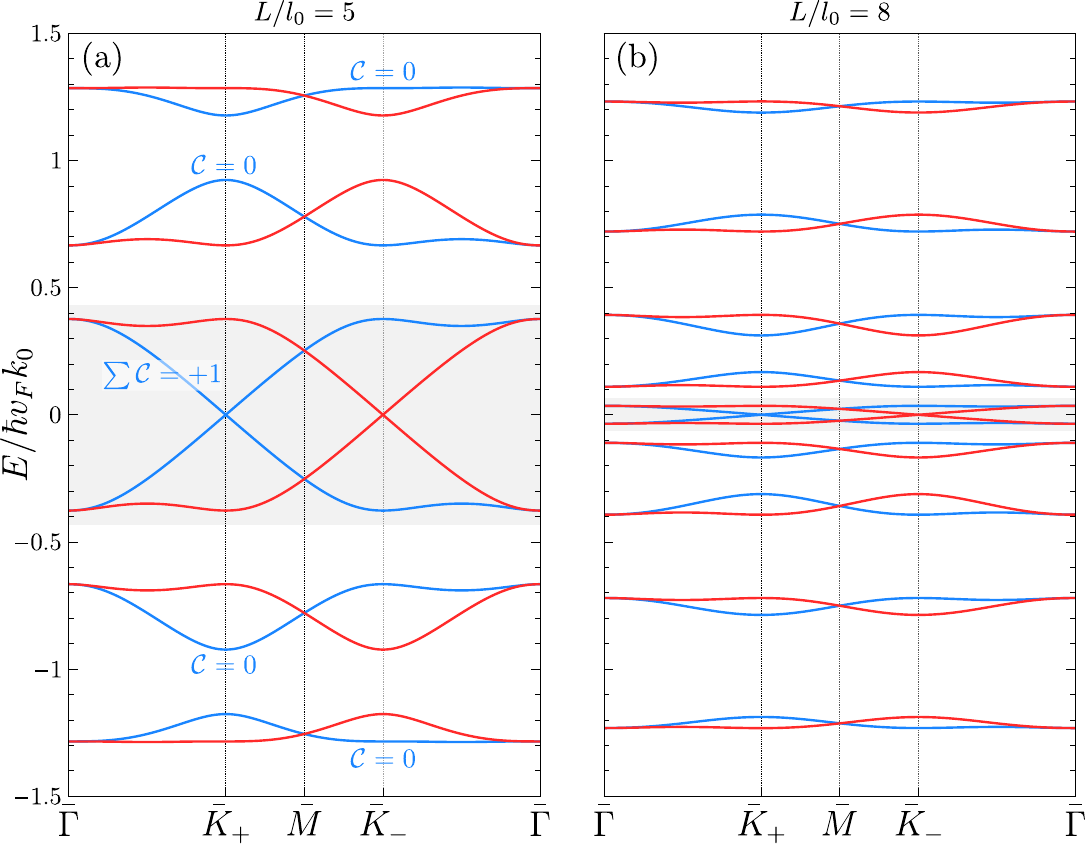}
    \caption{Band structure for valley $K_+$ (blue) and $K_-$ (red) of the valley-projected theory along high-symmetry lines of the SBZ [green path in Fig.\ \ref{fig:system}(c)] for (a) $L/l_0 = 5$ and (b) $L/l_0 = 8$, where energy is given in units $\hbar v_F k_0$ with $k_0 = 4\pi/3L$. The valley Chern numbers for $K_+$ are shown in (a) where the two nontrival bands are contained in the gray energy window.}
\label{fig:cbands}
\end{figure}

The resulting band structure is shown in Fig.\ \ref{fig:cbands} along high-symmetry lines of the SBZ for two realistic values of $L/l_0 > 1$, which is our regime of interest. Owing to the chiral symmetry, the bands are symmetric about zero energy and we label them as $E_{\nu,n}(\bm k) = \sgn(n) E_{|n|}(\nu \bm k)$ with $n$ a nonzero integer. Moreover, one can show that in the presence of chiral symmetry, the bands for a given valley always degenerate in pairs at zero energy \cite{Aharonov1979, Phong2022}. However, this crossing can be avoided by symmetry-allowed terms, which we assume are small as outlined in the previous paragraph. Since $\mathcal C_{2z}$ is broken and the valley projection implicitly breaks $\mathcal T$, the bands can have a nonzero Chern number. We find that the two bands near charge neutrality carry a net (valley) Chern number \cite{Phong2022},
\begin{equation}
    \mathcal C_{\nu,1} + \mathcal C_{\nu,-1} = \nu,
\end{equation}
while all remote low-energy bands that are shown in Fig.\ \ref{fig:cbands} are Chern trivial. Note that the net Chern number of the low-energy bands forms an obstruction to an effective lattice model for the single-valley physics that retains only the low-energy degrees of freedom. The low-energy band manifold can thus be thought of as a single Landau level that is half filled at charge neutrality. This situation bears resemblance to the system of monolayer graphene in the presence of a constant, real magnetic field. In that case, the $n=0$ Landau levels  are also half-filled at charge neutrality. Invoking chiral symmetry, one can argue that the Hall conductivity must be the same in magnitude but opposite in sign in gaps related by $E \mapsto -E$. This leads to the non-trivial prediction that the gaps right above and below the $n=0$ Landau level must have Hall conductivity equal to $\pm \frac{1}{2} \frac{e^2}{h}$ per valley and spin \cite{GS05, PGC06, ZTS05}. This half-integer quantum Hall effect stems from the $\pi$-quantized Berry phase of the Dirac cones, and is related to the parity anomaly \cite{H88, S91, FV12, L19}.  Anomaly cancellation is obtained when both valleys are simultaneously considered.

\begin{figure}
    \centering
    \includegraphics[width=\linewidth]{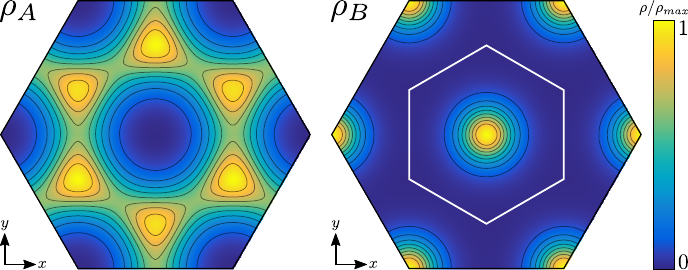}
    \caption{Density of the lowest conduction band ($n=1$) for $L/l_0 = 5$ [energy bands shown in Fig.\ \ref{fig:cbands}(a)] for sublattice $A$ ($\rho_A$, left) and $B$ ($\rho_B$, right) where the white hexagon gives the unit cell. The color scale is relative to the maximum of the density which is a factor $3$ larger for sublattice $B$.}
    \label{fig:density}
\end{figure}

We also show the density $\rho_{A/B}(\bm r)$, which is defined in App.\ \ref{app:cont}, of the lowest conduction band $(n=1)$ in Fig.\ \ref{fig:density}. Here, we omit the valley index since the density is the same for both valleys due to time-reversal symmetry. We observe that the charge density on sublattice $B$ is localized with support on a triangular lattice corresponding to the maxima of the PMF, while on sublattice $A$ it is extended with support on the PMF minima which together form a honeycomb lattice. Note also that $\rho_A$ breaks $\mathcal C_{2z}$ symmetry. The separation of the density of the two sublattices can be understood in analogy to the zeroth Landau level of graphene in a constant magnetic field, whose wave function only has support on one of the two sublattices, depending on the sign of the magnetic field and the valley \cite{Castro2009}. For a PMF, the valleys experience an opposite field and thus the support is identical for both valleys.

\subsection{Chiral interface modes}

Next, we consider the nodal lines of the PMF, which are given here by disjoint closed loops that encircle the maxima of the PMF, as shown in Fig.\ \ref{fig:system}(b). In the limit $L \gg l_0$, we can define a local valley Chern number inside and outside of these loops. For example, if the chemical potential lies in between the $m$th and $(m+1)$th local Landau levels, the net local valley Chern number is given by $\sgn(B_\nu) (m+1/2)$. The valley Chern number thus changes by an integer when the PMF changes sign. This results in a triangular lattice of circular domain walls given by the nodal lines of the PMF, that host well-defined chiral modes in the limit $L \gg l_0$.
\begin{figure}
    \centering
    \includegraphics[width=\linewidth]{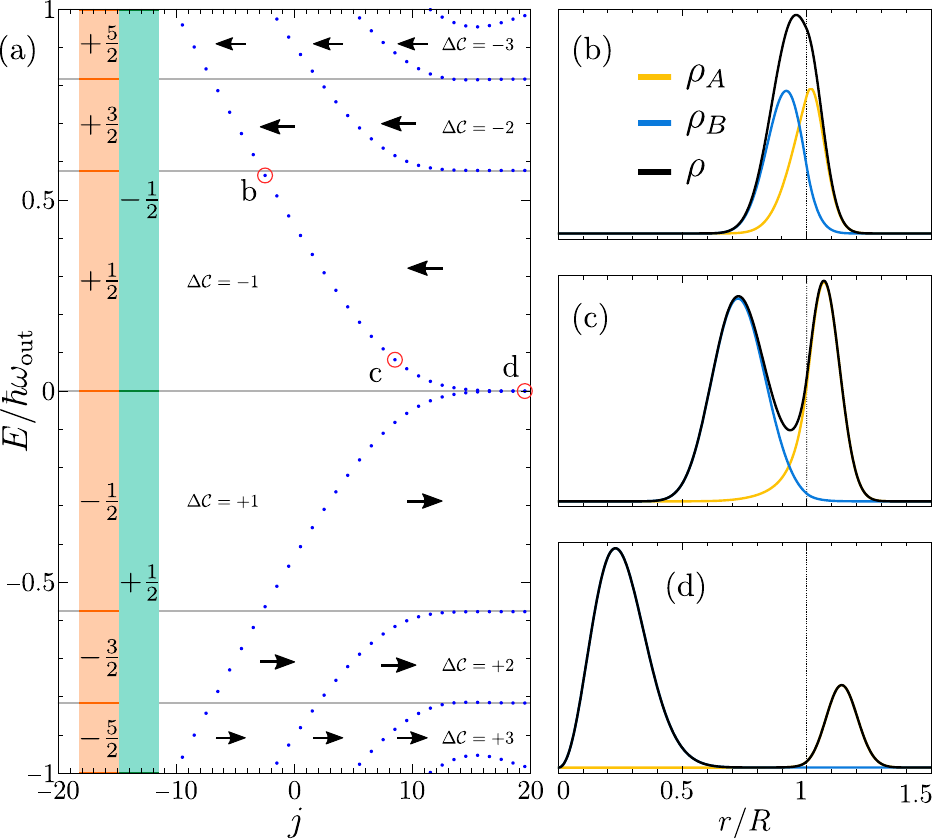}
    \caption{(a) Spectrum of a circular pseudomagnetic domain wall [Eq.\ \eqref{eq:mwall}] for valley $K_+$ where $\omega_\mathrm{out} = \sqrt{2} v_F/l_\mathrm{out}$. Orange and green columns show the local Landau level and Chern number sequence in the hexagons and triangles of the kagome tiling, respectively [see Fig.\ \ref{fig:system}(d)] and arrows indicate the number and orientation of domain-wall modes. (b-d) Single-particle density (arb.\ units) for states indicated in (a).}
    \label{fig:mstep}
\end{figure}

We demonstrate the existence of these domain-wall modes explicitly by considering an isolated nodal loop which we approximate by a circle. We then consider a piecewise constant PMF, $\bm B_\nu(\bm r) = \nu B(r) \bm e_z$ in polar coordinates $(r,\phi)$ where
\begin{equation} \label{eq:mwall}
B(r) = B_\mathrm{in} \Theta(R-r) + B_\mathrm{out} \Theta(r-R),
\end{equation}
with radius $R$. For concreteness, we take $R/l_\mathrm{in} = 6.5$ where $l_\mathrm{in} = \sqrt{\hbar/eB_\mathrm{in}}$ and $B_\mathrm{out} / B_\mathrm{in} = -3$ with $B_\mathrm{in} > 0$. We use the gauge $\bm A_\nu(\bm r) = \nu \left( r^2 - R^2 \right) B(r) \bm e_\phi/ 2 r$ with the spinor \emph{ansatz} $\Psi(\bm r) = e^{i\left(j-\nu \sigma_z/2 \right)\phi} \psi(r)$ where $j = \pm 1/2,\pm 3/2,\ldots$ is the eigenvalue of $J_z = -i\partial_\phi + \nu \sigma_z/2$. A normalizable solution can then be found in each region and the spectrum is obtained from continuity of the spinor at $r=R$. In Fig.\ \ref{fig:mstep}(a), we show the spectrum for valley $\nu=1$ versus $j$. Note that the net number of chiral modes matches the difference in the local valley Chern number between the inner and outer regions. In Fig.\ \ref{fig:mstep}(b), we show the single-particle density for different states. As before, $\rho_A$ ($\rho_B$) is mostly localized in the region with negative (positive) field. Moreover, for large $|j|$, the energy converges to the local Landau levels and the center of the single-particle wave function in the outer region moves away from the domain wall. Hence, the chiral modes percolate to neighboring loops at zero energy. This is illustrated in Fig.\ \ref{fig:system}(b), where the effective scattering centers form a kagome lattice.

For the long-wavelength physics, we can further replace the smooth PMF by an effective piecewise constant PMF with the same symmetry and connectivity. This yields a kagome tiling, as shown in Fig.\ \ref{fig:system}(d). The values of the constant pseudomagnetic field in the hexagons and triangles of the kagome tiling are determined by the condition that the net flux should vanish $\Phi_h + 2 \Phi_t = 0$ which gives $B_t = -3B_h$. The sequence of Chern numbers and interface modes for the kagome tiling are shown in Fig.\ \ref{fig:mstep}(a). The low-energy physics can thus be understood in terms of chiral modes bound to nodal lines of the PMF that percolate at charge neutrality with the connectivity of the oriented kagome network.

\section{Network model} \label{sec:network}

In the previous section, we argued that the single-valley low-energy physics of periodically strained graphene in the limit $L \gg l_0$ can be understood in terms of a kagome network of chiral modes; see Fig.\ \ref{fig:system}(d). Moreover, at low energies, each link of the network hosts a single chiral mode for a given valley and spin. In this section, we develop a phenomenological theory for the chiral kagome network. To this end, we assume that all microscopic details of the scattering process can be absorbed into an $S$ matrix that relates incoming and outgoing amplitudes at a point-like scattering node. Hence, we assume that the extent $l_0$ of the scattering region, which is of the order of the extent of the wavefunction of the chiral modes, is small compared to $L$. The $S$ matrix is only constrained by unitarity and the valley-preserving symmetries: $\mathcal C_{3z}$ and $\mathcal M_x \mathcal T$.
    \begin{figure}
    \centering
    \includegraphics[width=\linewidth]{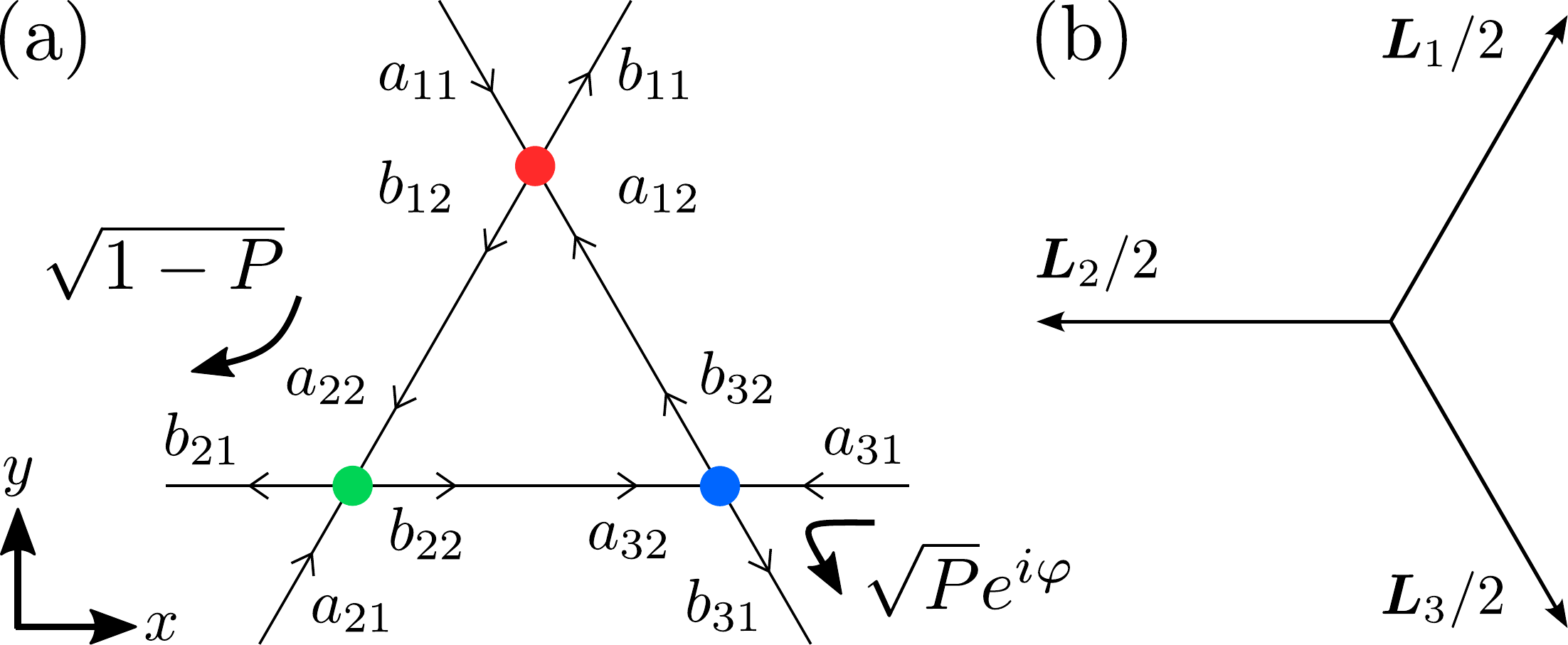}
    \caption{(a) Unit cell of the chiral kagome network. Here, arrows represent chiral modes propagating along oriented links and the three colored dots correspond to the three subnodes. (b) The lattice vectors $\bm L_1$, $\bm L_2$, and $\bm L_3$.}
    \label{fig:ncell}
\end{figure}
\begin{figure*}
    \centering
    \includegraphics[width=\linewidth]{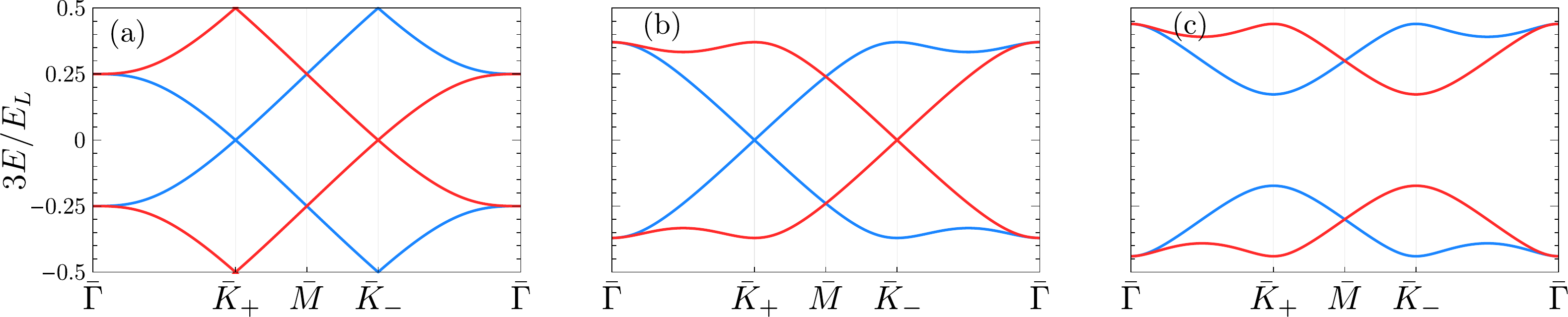}
    \caption{Network energy bands for valley $K_+$ (blue) and $K_-$ (red) in the fundamental domain along high-symmetry lines of the SBZ [green path in Fig.\ \ref{fig:system}(c)] for constant scattering parameters $(\varphi,P)$ as indicated on Fig.\ \ref{fig:phase}.}
    \label{fig:nbands1}
\end{figure*}

\subsection{Scattering matrix}

The unit cell of the chiral kagome network is shown in Fig.\ \ref{fig:ncell}(a). It consists of three scattering nodes which we refer to as \emph{subnodes}. Each cell can be labeled by a superlattice lattice vector $\bm R = n_1 \bm L_1 + n_2 \bm L_2$ with $n_{1,2}$ integers. Since the $S$ matrix only relates local incoming and outgoing scattering amplitudes, the total $S$ matrix of the unit cell is block diagonal in the subnodes:
\begin{equation} \label{eq:genS}
    \mathcal S = \begin{pmatrix} S_1 & 0 & 0 \\ 0 & S_2 & 0 \\ 0 & 0 & S_3 \end{pmatrix},
\end{equation}
such that $b_{\bm R} = \mathcal S a_{\bm R}$ where
\begin{align}
    a & = \left( a_{11}, a_{12}, a_{21}, a_{22}, a_{31}, a_{32} \right)^t, \\
    b & = \left( b_{11}, b_{12}, b_{21}, b_{22}, b_{31}, b_{32} \right)^t,
\end{align}
are the incoming and outgoing amplitudes of one cell, respectively, and which are defined in Fig.\ \ref{fig:ncell}. Threefold rotation symmetry $\mathcal C_{3z}$ yields $S_0=S_1=S_2=S_3$ and $\mathcal M_x \mathcal T$ gives $S_0 = ( S_0 )^t$. We thus have
\begin{equation} \label{eq:S0}
    S_0 = e^{iE_0} \begin{pmatrix} e^{i\varphi} \sqrt{P} & \sqrt{1-P} \\ \sqrt{1-P} & -e^{-i\varphi} \sqrt{P} \end{pmatrix},
\end{equation}
where $0 \leq P \leq 1$, and $\varphi$ and $E_0$ are phases. Here, $P$ is the probability for scattering along the acute corner of the triangle and $\varphi$ is the relative phase difference of the amplitudes, as illustrated in Fig.\ \ref{fig:ncell}(a). A detailed derivation of the $S$ matrix is given in App.\ \ref{app:network}.

\subsection{Link phases}

Neighboring scattering nodes in the network are connected by oriented links of length $L/2$ along which chiral modes freely propagate. During propagation between adjacent nodes, they acquire a phase $\lambda = \exp \left( -i2\pi E/E_L \right)$ with $E_L = 2hv / L$. Here, we assume that the modes have linear dispersion with slope $v \sim v_F$. Explicitly,
\begin{align}
    a_{11\bm R} & = \lambda b_{31\bm R-\bm L_3}, \label{eq:conn1} \\
    a_{12\bm R} & = \lambda b_{32\bm R}, \\
    a_{21\bm R} & = \lambda b_{11\bm R-\bm L_1}, \\
    a_{22\bm R} & = \lambda b_{12\bm R}, \\
    a_{31\bm R} & = \lambda b_{21\bm R-\bm L_2}, \\
    a_{32\bm R} & = \lambda b_{22\bm R}. \label{eq:conn2}
\end{align}
Since the network has translation symmetry, we Fourier transform to momentum space,
\begin{equation} \label{eq:Fourier}
    b_{\tau j\bm R} = \frac{1}{\sqrt{N}} \sum_{\bm k} e^{i\bm k \cdot \bm R} b_{\tau j\bm k},
\end{equation}
where $N$ is the number of cells, and similarly for incoming amplitudes. Note that we work in Bloch form. Here, $\tau=1,2,3$ is the subnode index, and $j=1,2$ is the link index. We then obtain
\begin{equation} \label{eq:link}
    a_{\bm k} = \lambda \mathcal M_{\bm k} b_{\bm k},
\end{equation}
where
\begin{equation}
    \mathcal M_{\bm k} = \begin{pmatrix}
    0 & 0 & 0 & 0 & e^{-ik_3} & 0 \\
    0 & 0 & 0 & 0 & 0 & 1 \\
    e^{-ik_1} & 0 & 0 & 0 & 0 & 0 \\
    0 & 1 & 0 & 0 & 0 & 0 \\
    0 & 0 & e^{-ik_2} & 0 & 0 & 0 \\
    0 & 0 & 0 & 1 & 0 & 0
    \end{pmatrix},
\end{equation}
is the connectivity matrix with $k_i = \bm k \cdot \bm L_i$ ($i=1,2,3$) and $\bm L_3 = - ( \bm L_1 + \bm L_2)$; see Fig.\ \ref{fig:ncell}(b). Since the links always connect different subnodes, $\mathcal M_{\bm k}$ is block antidiagonal.

\subsection{Network bands}

Combining $b_{\bm k} = \mathcal S a_{\bm k}$ with \eqref{eq:link} yields
\begin{equation} \label{eq:network}
    \mathcal U_{\bm k} b_{\bm k} = e^{i2 \pi E / E_L} b_{\bm k},
\end{equation}
where $\mathcal U_{\bm k} = \mathcal S \mathcal M_{\bm k}$. Solving for the energy gives
\begin{equation} \label{eq:nbands}
    E_{n,s}(\bm k) = \frac{E_L}{3} \left[ n + s\, \frac{\arccos f(\bm k)}{2\pi} \right],
\end{equation}
with $s=\pm1$ and integer $n$, and where we set $E_0 = -\pi/6$ to obtain symmetric bands. The result for the other valley is obtained by letting $\bm k \rightarrow - \bm k$. Here, we have also defined
\begin{equation} \label{eq:f}
    f(\bm k) = \sqrt{P} \left[ \left( 1 - P \right) \sum_{i=1}^3 \sin \left( k_i + \varphi \right) + P \sin 3 \varphi \right].
\end{equation}
Note that shifting the origin to $\bar K_\pm$ is equivalent to sending $\varphi \rightarrow \varphi \pm 2\pi/3$ in Eq.\ \eqref{eq:f}, respectively, such that all unique cases are contained in $\varphi \in [-\pi/3,\pi/3[$.

When the scattering parameters $P$ and $\varphi$ are constant, the energy enters only in the link phases and Eq.\ \eqref{eq:network} is periodic in energy with period $E_L$. However, the network energy bands in Eq.\ \eqref{eq:nbands}, shown in Fig.\ \ref{fig:nbands1}, have a smaller period $E_L/3$. This can be understood from the phase-rotation symmetry \cite{Delplace2020,Delplace2017},
\begin{equation} \label{eq:prs}
    D \mathcal U_{\bm k} D^{-1} = e^{-i2\pi/3} \mathcal U_{\bm k},
\end{equation}
where
\begin{equation}
    D = \textrm{diag} \left( 1, e^{i4\pi/3}, e^{i2\pi/3} \right) \otimes \mathds 1_2,
\end{equation}
such that $D b_{n,s}$ is an eigenstate with energy $E_{n+1,s} = E_{n,s} + E_L/3$. Note that $D^3 = 1$, corresponding to one full period and $b_{n+3,s} = b_{n,s}$. Hence, we only need to consider two bands, called the fundamental domain, chosen here as $-E_L/6 < E \leq E_L/6$. We further note that \eqref{eq:prs} is a result of the cyclic structure of the network. It is independent of symmetry constraints, and holds for any subnode block diagonal $S$ matrix.
\begin{figure}
    \centering
    \includegraphics[width=.95\linewidth]{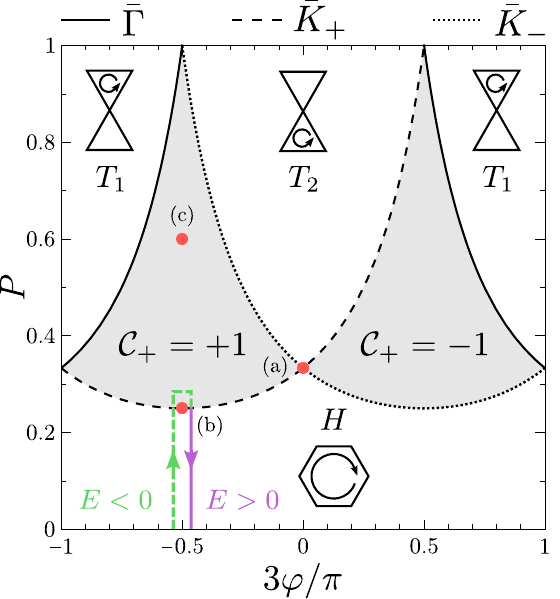}
    \caption{Phase diagram of the chiral kagome network. White (gray) regions are trivial (Chern) gapped phases, where the Chern number $\mathcal C_+$ of the band $E_{n,+}$ is shown. The curves correspond to gap closings at high-symmetry points and the dots indicate the parameters for the bands shown in Fig.\ \ref{fig:nbands1}.}
    \label{fig:phase}
\end{figure}

\subsection{Phase diagram}

The phase diagram of the oriented kagome network with a single channel has been discussed previously in the context of photonic systems \cite{Pasek2014}. Here, we discuss the phase diagram from a different perspective and give some new analytical results.

We start by identifying the gap closings between pairs of bands $E_{n,\pm}$ and between $E_{n,+}$ and $E_{n+1,-}$. In general, the gap closes for $f(\bm k)^2= 1$ [see Eq.\ \eqref{eq:nbands}] which occurs only at the high-symmetry points $\bar \Gamma$ and $\bar K_\pm$. This yields
\begin{equation}
    P_{\bar \Gamma} = \frac{1}{4 \sin^2 \varphi}, \qquad P_{\bar K_\pm} = \frac{1}{4 \sin^2 \left( \varphi \pm 2\pi/3 \right)},
\end{equation}
which define gap-closing lines in the $(\varphi,P)$ plane, as shown in Fig.\ \ref{fig:phase}. 

\subsubsection{Flatband limits}

Now consider the two flatband limits of the chiral kagome network, namely $P=0$ and $P=1$. In these limits, the bands are isolated and correspond to classical loop configurations, such that the Chern number vanishes. This can also be understood from a strong phase-rotation symmetry in the flatband limit, which reduces the fundamental domain to a single band \cite{Delplace2020,Delplace2017}. Since the Chern number is invariant under unitary transformations, it follows together with completeness that the sum of the Chern numbers in each fundamental domain vanishes. At a strong-phase rotation symmetric point, all bands are therefore Chern trivial. In the following, we refer to the phases that are adiabatically connected to the $P=0$ and $P=1$ flatband limit as the $H$ and $T$ phase, respectively. Indeed, for $P=0$, the flatbands correspond to hexagonal closed orbits in real space, while for $P=1$, the network is localized in triangular orbits, as illustrated in Fig.\ \ref{fig:phase}. As such, the two flatband limits correspond to two distinct classical limits. We further note that there are two $T$ phases, depending on which of the two triangles of the kagome network the density of a given band is mostly localized; see App.\ \ref{app:network}. The $H$ and $T$ phase are connected by a metallic phase at two points in the phase diagram where two gap-closing lines cross, called a percolation point. Hence, an interface between the trivial $H$ and $T$ phases supports a chiral mode which propagates in a snakelike fashion. This indicates that the Chern trivial phases do posses a notion of relative topology, similar to a massive Dirac electron subjected to a mass inversion on a domain wall.

\subsubsection{Chern phases and winding number}

For the remaining two gapped phases, we numerically calculate the Chern number \cite{Fukui2005} and we find two Chern phases; see Fig.\ \ref{fig:phase}, where pairs of bands $E_{n,\pm}$ carry opposite Chern numbers $\mathcal C_+ + \mathcal C_- = 0$ with $\mathcal C_+ = \pm 1$, as expected from phase-rotation symmetry. However, because of the unbounded spectrum of the network model which repeats periodically in energy, the total Chern number of the ground state is an ill-defined quantity. Indeed, there are two distinct gaps whose Chern number oscillates between zero and $\pm 1$. By calculating the spectrum in a ribbon geometry, we find that the boundary conditions pick out one particular sequence. 

We therefore need an additional invariant that is sensitive to the choice of unit cell. To this end, one can map the network to a piecewise constant Floquet lattice model \cite{Delplace2017,Delplace2020,DeBeule2020b}. The oriented kagome network can be mapped to a three-step Floquet on a honeycomb lattice \cite{Kitagawa2010,Delplace2020} but the details are beyond the scope of this work. For explicit examples, we refer to Refs.\ \onlinecite{Delplace2020,DeBeule2020b}. Importantly, this mapping is not unique and depends on the cell choice. The resulting effective Floquet models give rise to a continous unitary time evolution $U(t,\bm k)$ from $t=0$ to $t=T$ with $U(0,\bm k)=1$. In the flatband limit, $[0,T] \times \mathrm{BZ} \mapsto U$ is a periodic map and one can define a single Floquet winding number (see definition below) \cite{DeBeule2020b}. In general, however, the unitary is periodized by defining an effective Hamiltonian \cite{Rudner2014,Delplace2017}
\begin{equation} \label{eq:Heff}
    H^\mathrm{eff}_\eta(\bm k) = \frac{1}{iT} \log_{\eta} U(T,\bm k),
\end{equation}
where we assume that $U(T,\bm k)$ is gapped on the unit circle and $\eta$ is an angle inside a given gap. The effective Hamiltonian is not unique and depends on the branch cut of the logarithm, which we place inside the gap defined by $\eta$. Note that Eq.\ \eqref{eq:Heff} is only well-defined if we can place the branch cut inside a gap. The periodized unitary is defined as \cite{Maczewsky2017,Delplace2017}
\begin{equation}
    V_\eta(t,\bm k) = U(t,\bm k) e^{-i t H^\mathrm{eff}_\eta(\bm k)},
\end{equation}
such that $V_\eta(0,\bm k) = V_\eta(T,\bm k) = 1$ and which can be continuously deformed \cite{Rudner2014,CARPENTIER2015779} to $U(t,\bm k)$ without closing the gap at $\eta$. The bulk topological invariant is then given by the winding number of the periodic map $[0,T] \times \mathrm{BZ} \mapsto V_\eta$
\begin{equation}
    \begin{aligned}
        W_\eta[U] & = \frac{1}{8\pi^2} \int_0^T dt \int_\mathrm{BZ} d^2\bm k \\
        & \mathrm{Tr} \left\{ V_\eta^{-1} \left( \partial_t V_\eta \right) \left[ V_\eta^{-1} \left( \partial_{k_x} V_\eta \right), V_\eta^{-1} \left( \partial_{k_y} V_\eta \right) \right] \right\}.
    \end{aligned}
\end{equation}
The Floquet winding number $W$ is thus defined for each gap of the Floquet quasispectrum and a difference in winding numbers of two gaps equals the net Chern number of the intervening bands \cite{Maczewsky2017}. Moreover, the winding number obeys the bulk-boundary correspondence \cite{Rudner2014}
\begin{equation}
    n_\mathrm{edge}(\eta) = W_\eta. 
\end{equation}
where $n_\mathrm{edge}(\eta)$ is the net number (accounting for chirality) of chiral edge modes in the gap at $\eta$.

We conclude that a scattering network is characterized by several winding numbers \cite{Delplace2017}, one for each unique choice of unit cell with a corresponding set of edge types. Moreover, when the network is in a Chern trivial phase, the winding number is the same for each gap, while it alternates in a Chern phase. We can understand the need for multiple winding numbers intuitively as follows. Certain boundary configurations of the network have no counterpart in the effective Floquet theory for a given choice of cell. For example, in the kagome network, triangle and hexagon edges are realized by two different Floquet models. In addition to the Chern number, the chiral kagome network is thus also characterized by winding numbers $(W_I,W_{II})$ for each gap, where $W_I$ corresponds to our choice of unit cell shown in Fig.\ \ref{fig:ncell}(a), and $W_{II}$ corresponds to a cell whose subnodes all belong to a single hexagon of the kagome network.
\begin{figure}
    \centering
    \includegraphics[width=\linewidth]{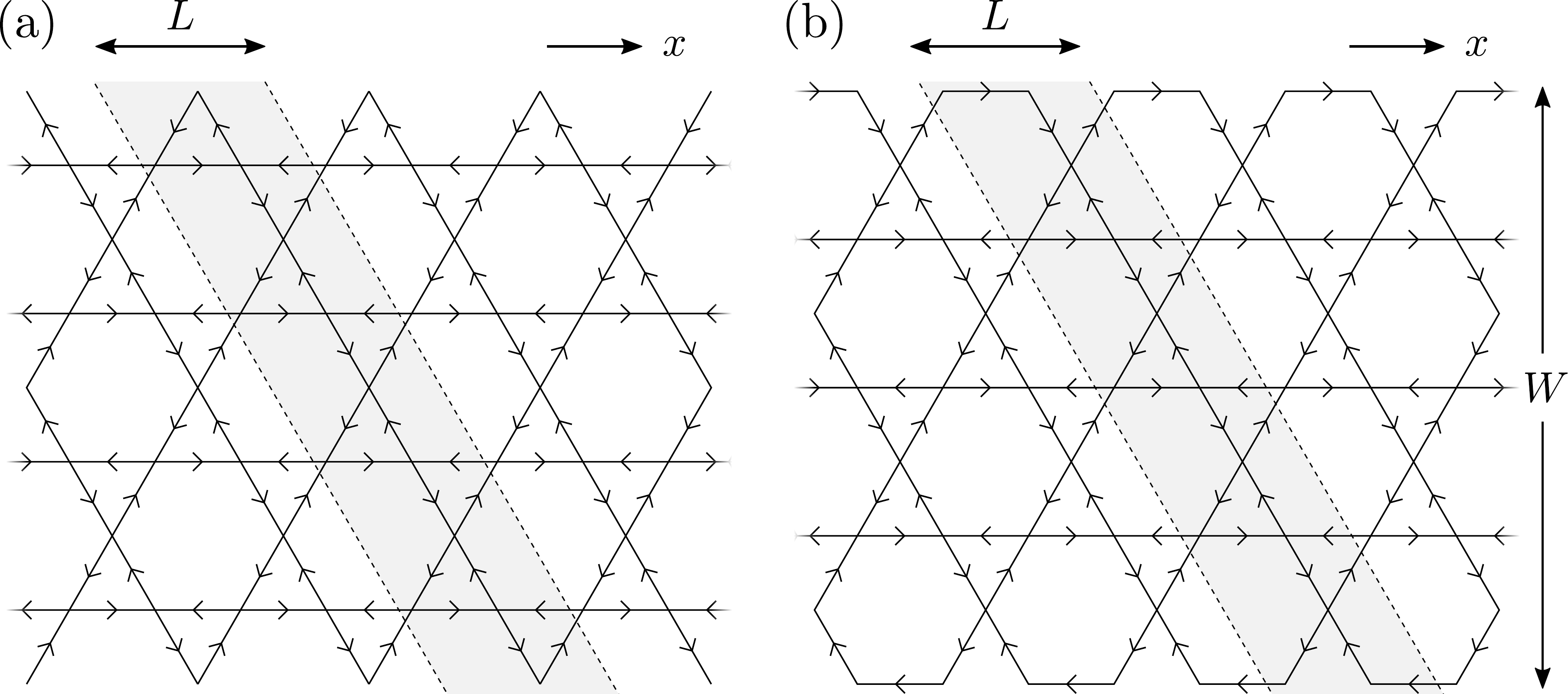}
    \caption{Network ribbons of width $M=2$ with (a) triangle and (b) hexagon edges. The shaded region gives the unit cell.}
    \label{fig:edges}
\end{figure}
\begin{figure*}
    \centering
    \includegraphics[width=\linewidth]{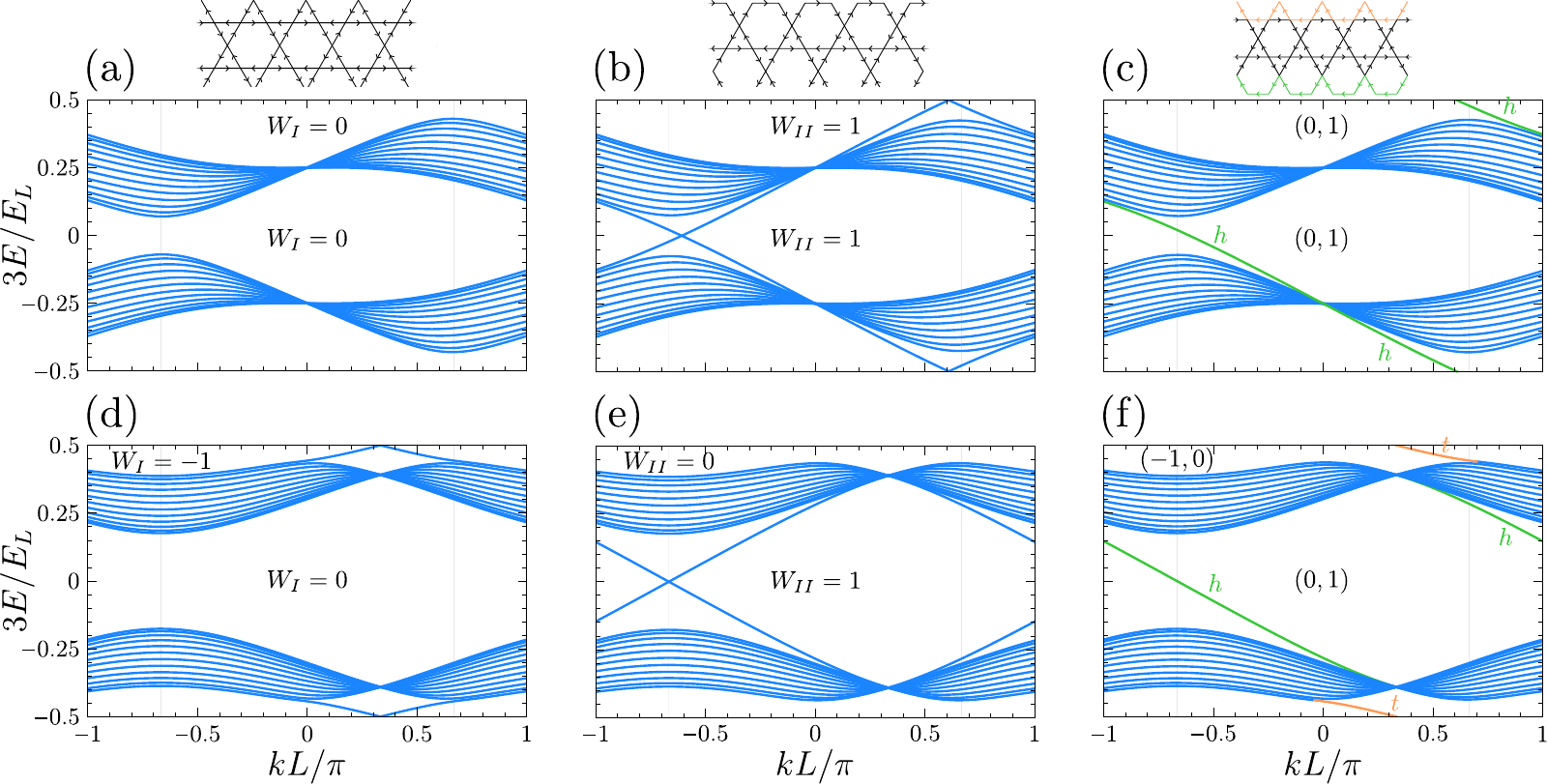}
    \caption{(a--c) Energy bands of a network ribbon with triangle (a), hexagon (b), and mixed (c) edges of width $M=12$ for valley $\nu=1$ in the $T$ phase [$(\varphi,P)=(0,1/2)$]. Here, vertical gray lines indicate the projected $\bar K_\pm$ points at $k = \mp 2\pi/3L$ and the relevant winding numbers $(W_I,W_{II})$ of the gaps are are indicated. (d--f) Same for the $\mathcal C_+=1$ phase [(c) in Fig.\ \ref{fig:phase}].}
    \label{fig:ribbonbands}
\end{figure*}

\subsection{Network ribbon}

We now consider a ribbon periodic along the $x$ direction with period $L$ and finite along the $y$ direction with width $W = \sqrt{3} \, M L$ where $M=1,2,\ldots$. Moreover, we consider two types of boundaries: triangle and hexagon edges, illustrated in Fig.\ \ref{fig:edges}(a) and (b), respectively. These are the simplest edge types that respect current conservation, i.e., at each scattering node the total number of incoming modes should equal the total number of outgoing modes. The energy bands for ribbons with triangle and hexagon edges are shown in Figs.\ \ref{fig:ribbonbands}(a,b) and (d,e) for the trivial $T$ and Chern phase, respectively. In the trivial phases, the existence of edge modes depends on the type of edge. Indeed, in the flatband limit of the $T$ phase ($P=1$), a hexagon edge is decoupled from the bulk which is localized in triangular orbits. Hence, the $T$ phase supports edge modes at a hexagon edge while there are no edge modes at a triangle edge. The opposite situation then holds for the $H$ phase. In the Chern phases, edge modes exist at any boundary. However, as we mentioned previously, while differences of Chern numbers are well-defined, the net Chern number is undetermined because of the unbounded spectrum. Moreover, the net Chern number depends in principle on the entire set of occupied bands. This includes bands that are not captured by the network model, which is only valid in a finite energy window near charge neutrality. It is thus a priori not clear which of the two gaps in the fundamental domain hosts edge modes. This is resolved in the network model by accounting for the winding numbers $(W_I,W_{II})$ as discussed in the previous section. Here, we use the bulk-boundary correspondence \cite{Rudner2014} to obtain the winding numbers from the finite-size calculation; see Fig.\ \ref{fig:ribbonbands}.

Finally, we consider a ribbon with a hexagonal edge and a triangle edge on opposite sides. In this case, each trivial phase hosts edge modes but only on the edge that does not support a localized bulk mode. At the topological phase transition, the edge mode moves to the opposite edge without changing propagation direction. This is consistent with the fact that the hexagonal and triangle edge naturally support edge modes of opposite chirality (see Fig.\ \ref{fig:edges}). We find single-boundary edge modes in all gapped phases for ribbons with both types of edges. The mixed-edge ribbon is thus characterized by three invariants: the Chern number of one of the two bands in the fundamental domain and a pair of winding numbers, one for each edge. The corresponding spectrum is shown in Fig.\ \ref{fig:ribbonbands}(c,f), where we have indicated the boundary for each edge mode. In a rectangular sample, it is not immediately clear what happens to the single-boundary edge modes. We find that in a theory with local current conservation, such a setup is only possible if the system is coupled to a reservoir.
\begin{figure}
    \centering
    \includegraphics[width=0.85\linewidth]{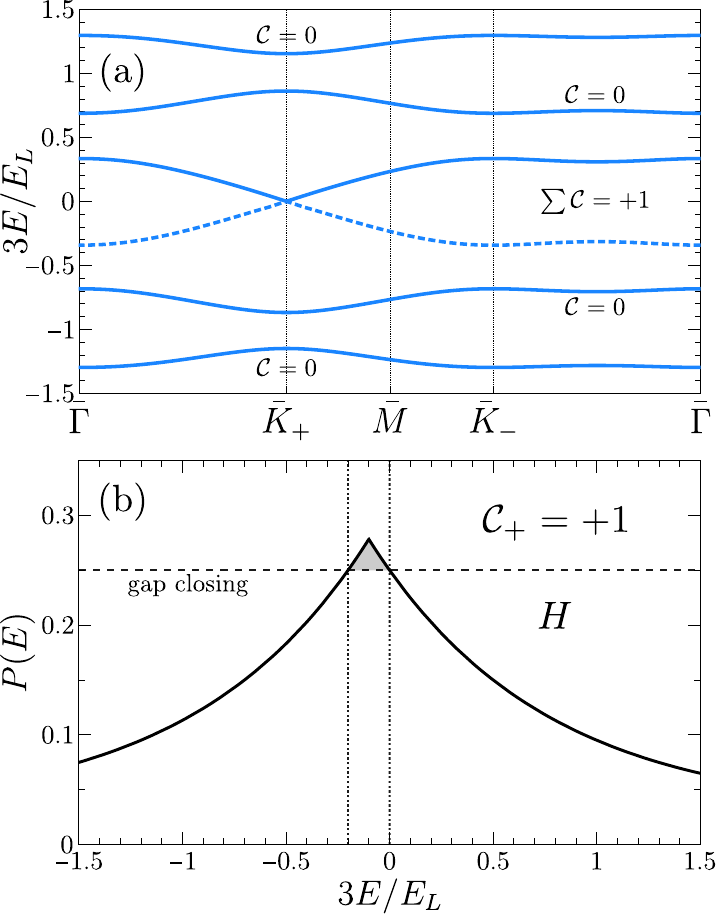}
    \caption{(a) Network energy bands for valley $K_+$ along high-symmetry lines of the SBZ [green path in Fig.\ \ref{fig:system}(c)] for $\varphi = -\pi/6$ and $P(E)$ as shown in (b). The corresponding path in the $(\varphi,P)$ plane is shown in Fig.\ \ref{fig:phase} as the green dashed ($E<0$) and solid purple ($E>0$) line. The dashed band carries unit Chern number, while all other bands are Chern trivial. In (b), the white (shaded) region below the curve corresponds to the $H$ (Chern) phase of the chiral kagome network.}
    \label{fig:nbands2}
\end{figure}

\section{Discussion} \label{sec:disc}

From our discussion on the wave functions of the circular domain wall, shown in Fig.\ \ref{fig:mstep}(b), it is clear that the $S$ matrix should depend on the energy. As we approach zero energy, the chiral domain-wall modes merge with the bulk Landau level and the wavefunction separates into two parts with exponentially suppressed overlaps. Hence, the network modes of neighboring nodal loops strongly overlap near charge neutrality. At zero energy, the network therefore undergoes a percolation transition, similar to a quantum Hall transition at critical filling \cite{Chalker1988,Kramer2005,Potter2020}. This implies that the $S$ matrix traces out a path in the $(\varphi,P)$ plane as a function of energy and Eq.\ \eqref{eq:network} therefore becomes a nonlinear eigenvalue problem that is solved self-consistently. Percolation then corresponds to crossing phase boundary where the gap closes. Further note that the orientation of the network is reversed as the energy changes sign, since this reverses the difference in the local valley Chern number. Reversing the network orientation amounts to letting $\mathcal M_{\bm k} \rightarrow \left( \mathds 1_3 \otimes \sigma_x \right) \mathcal M_{-\bm k} \left( \mathds 1_3 \otimes \sigma_x \right)$ in the network model.

Ideally, one obtains the energy-dependence of the $S$ matrix from a microscopic theory. Here, we only demonstrate that we can qualitatively reproduce both the energy bands and their topology by imposing an appropriate energy dependence. To this end, we require that the resulting bands are approximately symmetric and gapless at zero energy. Moreover, the net valley Chern number of the two bands near charge neutrality should equal $\nu=\pm1$. These conditions are all met in Fig.\ \ref{fig:nbands2}(a) by taking scattering parameters $\varphi=-\pi/6$ and $P=P(E)$ as shown in Fig.\ \ref{fig:nbands2}(b). The corresponding path in the $(\varphi,P)$ plane is shown in the bottom left corner of Fig.\ \ref{fig:phase} where the path for positive (negative) energies is shown as the solid (dashed) line. We offset the paths from $\varphi=-\pi/6$ in the figure for clarity. This path is chosen such that part of the highest valence band dips into the Chern phase, picking up a unit Chern number. We note that this energy dependence results in a tiny energy gap around zero energy which allows us to calculate the Chern number numerically \cite{Fukui2005}. This gap can be made arbitrarily small by decreasing the energy window of the Chern phase; see Fig.\ \ref{fig:nbands2}(b). 
The energy bands thus obtained are in qualitative agreement with those obtained from the continuum model. However, the energy scale in Fig.\ \ref{fig:nbands2}(a) is different than that in Fig.\ \ref{fig:cbands} by a factor $v/v_F$ which controls the bandwidth. Hence, the average group velocity $v$ of the chiral network modes is smaller than the Fermi velocity of graphene. In reality, the group velocity should also depend on the energy; see Fig.\ \ref{fig:mstep}(a).

While the network model can qualitatively capture both the energy bands and the topology of the bulk low-energy physics of the periodic PMF, it cannot address the boundary physics discussed in Ref.\ \onlinecite{Phong2022}. This is a consequence of the coarse-grained approach for the network that breaks down at the boundary. Indeed, the links of the network correspond to structures on the scale of the strain field, while all microscopics is smoothed over. Although both the network and atomistic models \cite{Phong2022} support single-boundary edge modes, there is no clear connection between these two cases. Indeed, there is no well-defined bulk-boundary correspondence when both valleys are taken into account because the total Chern number inevitably vanishes when $\mathcal T$ symmetry is restored. As such, the fate of the single-valley edge modes depends on intervalley scattering at the boundary.

In conclusion, we have demonstrated that the single-valley long-wavelength physics of periodically strained graphene can be qualitatively understood with an oriented network model. Here, the symmetries and connectivity of the network are solely determined by the nodal lines of the corresponding pseudomagnetic field which host chiral domain-wall modes. In this work, we considered a commensurate strain field with $C_{3v}$ symmetry which is relevant for graphene on NbSe$_2$. For this system, the network corresponds to a chiral kagome network constrained by the valley-preserving symmetries of the pseudomagnetic field. Moreover, the presence of a low-energy valley chiral network is only dependent on the nodal structure of the pseudomagnetic field, as long as the pseudomagnetic varies slowly with respect to the effective magnetic length. Hence, network physics should be ubiquitous in strained graphenes and robust against deviations from perfect periodicity. We finally note that, compared to atomistic methods, the network model is computationally cheap and can therefore be a valuable tool for investigating electronic transport close to charge neutrality in periodically strained graphene. In fact, a corollary of the presence of a low-energy electronic network is the potential for interference phenomena in a mesoscopic transport setup upon applying a weak real magnetic field normal to the graphene plane, or an in-plane electric field.

\begin{acknowledgments}
This research was funded in whole, or in part, by the Luxembourg National Research Fund (FNR) (project No.\ 16515716). Work by VTP and EJM is supported by the Department of Energy under grant DE-FG02-84ER45118. VTP acknowledges further support from the P.D. Soros Fellowship for New Americans and the National Science Foundation's Graduate Research Fellowships Program. For  the purpose of open access, the authors have applied a Creative Commons Attribution 4.0 International (CC BY 4.0) license to any Author Accepted Manuscript version arising from this submission.
\end{acknowledgments}

\appendix 

\section{Valley-projected theory} \label{app:cont}

\subsection{Continuum model}

Here, we present more details on the continuum Hamiltonian given in Eq.\ \eqref{eq:ham}. For simplicity, we take the gauge $\bm A_\nu = \nu A(\bm r) \bm e_x$ where
\begin{equation}
    A(\bm r) = -B_0\sum_{i=1}^3 \frac{\sin ( \bm G_i \cdot \bm r + \phi )}{G_{iy}},
\end{equation}
with
\begin{equation}
    \bm G_1 = \frac{4\pi}{\sqrt{3}L} \begin{pmatrix} 0 \\ 1 \end{pmatrix}, \qquad \bm G_{2,3} = \frac{4\pi}{\sqrt{3}L} \begin{pmatrix} \mp \sqrt{3}/2 \\ -1/2 \end{pmatrix}.
\end{equation}
Up to a gauge transformation, this gauge is equivalent to the gauge that is directly related to the strain field \cite{Vozmediano2010},
\begin{equation}
    \bm A_\nu(\bm r) = \frac{\nu \hbar}{e} \frac{\sqrt{3}\beta}{2a} \begin{pmatrix} \varepsilon_{xx} - \varepsilon_{yy} \\ -2\varepsilon_{xy} \end{pmatrix},
\end{equation}
where $\nu = \pm 1$ is the valley index, $\beta \sim 1$ a dimensionless material constant, $a \approx 0.25$~nm the graphene lattice constant, and $\varepsilon_{ij}$ the strain tensor. The corresponding pseudomagnetic field (PMF) is given in Eq.\ \eqref{eq:pmf} which has vanishing net flux,
\begin{equation}
    \frac{1}{V_c} \int_{cell} d^2 \bm r \, \bm B_\nu \cdot \bm e_z = \nu B_{\bm G=\bm 0} = 0,
\end{equation}
with $V_c = \sqrt{3} L^2 /2$ the area of the unit cell.

Remember that $L=Na$ with integer $N \gg 1$ such that the PMF varies slowly on the graphene lattice scale. In this case, the single-valley theory is a good approximation. We diagonalize the continuum Hamiltonian in Eq.\ \eqref{eq:ham}  by Fourier transform:
\begin{equation}
    \hat \psi_\nu (\bm r) = \frac{1}{\sqrt{V}} \sum_{\bm k \in \textrm{SBZ}} \sum_{\bm G} e^{i \left( \bm k + \bm G \right) \cdot \bm r} \hat c_{\nu \bm k + \bm G},
\end{equation}
where the sum over momentum has been written as a sum over the superlattice Brillouin zone (SBZ) and a sum over reciprocal lattice vectors of the superlattice. The valley-projected Hamiltonian given in Eq.\ \eqref{eq:ham} becomes
\begin{align}
    \hat H_\nu & = \hbar v_F \sum_{\bm k \in \textrm{SBZ}}  \sum_{\bm G, \bm G'}  \hat c_{\nu \bm k+\bm G}^\dag \Big[ \nonumber \\
    & + \delta_{\bm G, \bm G'} \left( \bm k + \bm G - \bar{\bm K}_\nu \right) \cdot \left( \nu \sigma_x, \sigma_y \right) \\
    & + \sum_{i=1}^3 \frac{e^{-i\phi} \delta_{\bm G + \bm G_i,\bm G'} - e^{i\phi} \delta_{\bm G - \bm G_i,\bm G'}}{2il_0^2G_{iy}} \, \sigma_x \Big] \hat c_{\nu \bm k + \bm G'}, \nonumber
\end{align}
where we have placed the momentum origin at $\bar{\bm K}_\nu = \nu \left( 4\pi/3L \right) \bm e_x$. Next, we truncate the Bloch Hamiltonian by only taking $|\bm G| < k_c$ where $k_c$ is increased until the energy bands under consideration converge. Because of chiral symmetry, the bands are symmetric about zero energy and we label them as $E_{\nu,n}(\bm k) = \sgn(n) E_{|n|}(\nu \bm k)$ with $n$ a nonzero integer. The corresponding Bloch wave function is written as $\psi_{n\bm k}(\bm r) = e^{i \left( \bm k - \bar{\bm K}_\nu \right) \cdot \bm r} u_{n\bm k}(\bm r) / \sqrt{N}$ with
\begin{equation} 
    u_{n\bm k}(\bm r) = \frac{1}{\sqrt{V_c}} \sum_{\bm G} e^{i\bm G \cdot \bm r} \phi_{n\bm k,\bm G},
\end{equation}
where $N$ is the number of cells and $\phi_{n\bm k}$ are eigenstates of the Bloch Hamiltonian that we obtain numerically. The total density of sublattice $\sigma = A,B$ for band $n$ is
\begin{equation}
    \rho_{n\sigma}(\bm r) = \frac{1}{N} \sum_{\bm k \in \textrm{SBZ}} u_{n\bm k}^\dag(\bm r) \left( \frac{\sigma_0 \pm \sigma_z}{2} \right) u_{n\bm k}(\bm r),
\end{equation}
with a positive $(+)$ sign for sublattice $A$ and a negative sign $(-)$ for sublattice $B$.

Throughout this paper, we work in dimensionless units by sending
\begin{align}
    \bm k & \rightarrow k_0 \bm k, \\
    E & \rightarrow \hbar v_F k_0 E,
\end{align}
with $k_0 = 4\pi/3L$. In these units, the Hamiltonian only contains one dimensionless parameter
\begin{equation}
    \left( k_0 l_0 \right)^{-1} = \frac{3}{4\pi} \frac{L}{l_0},
\end{equation}
where (for $t \approx 3$~eV)
\begin{align}
    \hbar v_F k_0 = \frac{2\pi t}{\sqrt{3} N} & \approx \frac{11 \, \textrm{eV}}{N}, \\
    \frac{L}{l_0} & \approx \frac{N}{100} \sqrt{\frac{B_0}{T}}.
\end{align}
For example, for $N=60$ ($L\approx 15$~nm) and $B_0 = 100$~T, which are realistic values for engineering a periodic strain field with a varying height profile \cite{Phong2022} of the order of $6$~\AA, we obtain $L/l_0 \approx 6$ and $\hbar v_F k_0 \approx 180$~meV.

\subsection{Symmetries}

Here, we show how the (pseudo) vector potential is constrained by the symmetries of the system. Note that the $\mathcal C_{3z}$ symmetry inherited from the strain field is not the microscopic $\mathcal C_{3z}$ symmetry, which has to be broken since otherwise the strain tensor is only given by its trace (pure dilatation) and the pseudovector potential would be absent. Indeed, one can think about the pseudovector potential in terms of a strain-induced local shift of the Dirac point from the zone corner, to which it would be fixed when microscopic $\mathcal C_{3z}$ were conserved. The remaining symmetries that we consider are spinless time reversal ($\mathcal T$), twofold rotation ($\mathcal C_{2z}$), and mirror symmetry across the $y$ axis $(\mathcal M_x)$. We take the following representation for the action of the symmetries on the field operators:
\begin{align}
    \hat{\mathcal T} \hat \psi_\nu (\bm r) \hat{\mathcal T}^{-1} & = \hat \psi_{-\nu} (\bm r) \\
    \hat{\mathcal C}_{2z} \psi_\nu (\bm r) \hat{\mathcal C}_{2z}^{-1} & = \sigma_x \hat \psi_{-\nu} (-\bm r) \\
    \hat{\mathcal M}_x \hat \psi_\nu (x,y) \hat{\mathcal M}_x^{-1} & = \hat \psi_{-\nu} (-x,y) \\
    \hat{\mathcal C}_{3z} \hat \psi_\nu (\bm r) \hat{\mathcal C}_{3z}^{-1} & = e^{i\nu(2\pi/3) \sigma_z} \hat \psi_{\nu} (C_{3z} \bm r).
\end{align}
Next, we investigate how these symmetries act on the (pesudo)vector potential. Under time reversal, we find from $\hat{\mathcal T} \hat H \hat{\mathcal T}^{-1} = \hat H$ with $\hat H = \hat H_+ + \hat H_-$,
\begin{equation}
    \mathcal T: \quad \bm A(\bm r) \rightarrow -\bm A(\bm r), \quad \bm A_\nu(\bm r) \rightarrow \bm A_\nu (\bm r),
\end{equation}
where $\bm A$ ($\bm A_\nu$) corresponds to a real vector potential (pseudo vector potential) and we used $\mathcal T i \mathcal T^{-1} = -i$. We see that a PMF conserves $\mathcal T$. On the other hand, under a twofold rotation, we have
\begin{equation}
\mathcal C_{2z}: \quad \bm A(\bm r) \rightarrow -\bm A(-\bm r), \quad
\bm A_\nu (\bm r) \rightarrow \bm A_\nu(-\bm r),
\end{equation}
such that $\mathcal C_{2z}$ is conserved when $\bm A(\bm r) = -\bm A(-\bm r) + \nabla f$ for a real magnetic field and $\bm A_\nu(\bm r) = \bm  A_\nu(-\bm r) + \nabla f$ for a PMF, where $f(\bm r)$ is a scalar function. Hence, the real (pseudo)magnetic field has to be an even (odd) function of the position vector. The PMF that we consider in this work [see Eq. \eqref{eq:pmf}] generally does not conserve $\mathcal C_{2z}$ which is in fact broken by the substrate. Only the special case $\phi = \pi/2$ conserves $\mathcal C_{2z}$, yielding a gapless spectrum that can be mapped to a triangular chiral network. Finally, the mirror $\mathcal M_x$ gives
\begin{align}
    \bm A(x,y) & \rightarrow \begin{bmatrix} -A_x(-x,y) \\ +A_y(-x,y) \end{bmatrix}, \\
    \bm A_\nu(x,y) & \rightarrow \begin{bmatrix} +A_{\nu x}(-x,y) \\ -A_{\nu y}(-x,y) \end{bmatrix},
\end{align}
such that for a PMF, $\mathcal M_x$ is conserved if
\begin{align}
    A_{\nu x}(x,y) & = +A_{\nu x}(-x,y) + \partial_x f, \\
    A_{\nu y}(x,y) & = -A_{\nu y}(-x,y) + \partial_y f,
\end{align}
and $\bm B_\nu(x,y) = \bm B_\nu(-x,y)$. Note that in general, the mirror axis is offset to $x=-2\phi/3$.

Finally, we note that the Hamiltonian in Eq.\ \eqref{eq:ham} also has a chiral symmetry,
\begin{equation}
    \hat{\mathcal C} \hat{\psi}_\nu (\bm r) \hat{\mathcal C}^{-1} = \hat \psi_{\nu}^\dag (\bm r)\sigma_z, \qquad \hat{\mathcal C} i \hat{\mathcal C}^{-1} = -i,
\end{equation}
such that $\hat{\mathcal C} \hat H_\nu \hat{\mathcal C}^{-1} = \hat H_\nu$. However, this is a model-dependent symmetry due to the absence of terms proportional to $\sigma_0$ or $\sigma_z$ in the Hamiltonian. Such terms are in principle allowed by symmetry, e.g., a strain-induced pseudo-electrostatic potential or a constant sublattice-staggering term, respectively. The latter is proportional to the trace of the strain tensor \cite{Suzuura2005,Low2011}. In the tight-binding framework, the pseudo-electrostatic potential originates from a modulation of intra-sublattice hopping terms.

\section{Network model} \label{app:network}

\subsection{Symmetry constraints}

Under the threefold rotation, the subnodes of the kagome network undergo a cyclic permutation $1\rightarrow2\rightarrow3\rightarrow1$ and therefore $S_0 = S_1 \sim S_2 \sim S_3$ where $\sim$ indicates that the equality holds up to a diagonal unitary transformation. However, the resulting phases can always be removed by choosing a suitable basis for the amplitudes, i.e., $\left( a_1, a_2, a_3 \right) \rightarrow \left(a_1, U_2 a_2, U_3 a_3 \right)$ and similar for outgoing amplitudes, where $U_2$ and $U_3$ are diagonal $2\times2$ unitary matrices.

The composite symmetry $\mathcal M_x \mathcal T$ is more tricky. First, we consider the action of $\mathcal M_x$ and $\mathcal T$ separately. To this end, we need to consider both valleys. For subnode $1$ [red dot in Fig.\ \ref{fig:ncell}(a)] we have
\begin{equation}
    \begin{pmatrix} b_1 \\ b_1' \end{pmatrix} = \begin{pmatrix} S_1 & 0 \\ 0 & S_1' \end{pmatrix} \begin{pmatrix} a_1 \\ a_1' \end{pmatrix},
\end{equation}
where the unprimed and primed quantities correspond to valleys $K_+$ and $K_-$, respectively. Here, we have assumed that there is no intervalley scattering. Under $\mathcal M_x$, the amplitudes transform as
\begin{align}
    & \mathcal M_x a_1 \mathcal M_x^{-1} = a_1', \\
    & \mathcal M_x b_1 \mathcal M_x^{-1} = b_1',
\end{align}
which is clear from Fig.\ \ref{fig:ncell}(a). We then obtain
\begin{equation}
    \begin{pmatrix} b_1' \\ b_1 \end{pmatrix} = \mathcal M_x \begin{pmatrix} S_1 & 0 \\ 0 & S_1' \end{pmatrix} \mathcal M_x^{-1} \begin{pmatrix} a_1' \\ a_1 \end{pmatrix},
\end{equation}
or
\begin{equation}
    \mathcal M_x \begin{pmatrix} S_1 & 0 \\ 0 & S_1' \end{pmatrix} \mathcal M_x^{-1} = \begin{pmatrix} S_1' & 0 \\ 0 & S_1 \end{pmatrix}.
\end{equation}
When $\mathcal M_x$ is a symmetry, we have $S_1 \sim S_1'$. On the other hand, time-reversal symmetry yields
\begin{align}
    & \mathcal T a_1 \mathcal T^{-1} = \left( b_1' \right)^*, \\
    & \mathcal T b_1 \mathcal T^{-1} = \left( a_1' \right)^*,
\end{align}
or
\begin{equation}
    \mathcal T \begin{pmatrix} S_1 & 0 \\ 0 & S_1' \end{pmatrix} \mathcal T^{-1} = \begin{pmatrix} S_1' & 0 \\ 0 & S_1 \end{pmatrix}^t,
\end{equation}
such that $S_1 \sim \left( S_1' \right)^t$ when $\mathcal T$ is conserved. The combined symmetry $\mathcal M_x \mathcal T$ yields $S_1 \sim \left( S_1 \right)^t$. Since the diagonal unitary only acts within subnode $1$, it can removed independently of the phases that were removed under the constraint given by $\mathcal C_{3z}$.

In summary, threefold rotation symmetry ($\mathcal C_{3z}$) and mirror symmetry combined with time-reversal symmetry ($\mathcal M_x \mathcal T$) constrains the $S$ matrix as follows:
\begin{equation}
    \mathcal S =  \mathds 1_3 \otimes S_0, \qquad S_0 = \left( S_0 \right)^t.
\end{equation}
Any symmetric unitary matrix can be written as $S_0 = \exp \left(i X \right)$ with $X=X^t$ real symmetric. In this case, we can write $X = E_0 \sigma_0 + d_1 \sigma_1 + d_3 \sigma_3$ with three real parameters, which is equivalent to Eq.\ \eqref{eq:S0}.

\subsection{Classical loop configurations}

\begin{figure}
    \centering
    \includegraphics[width=\linewidth]{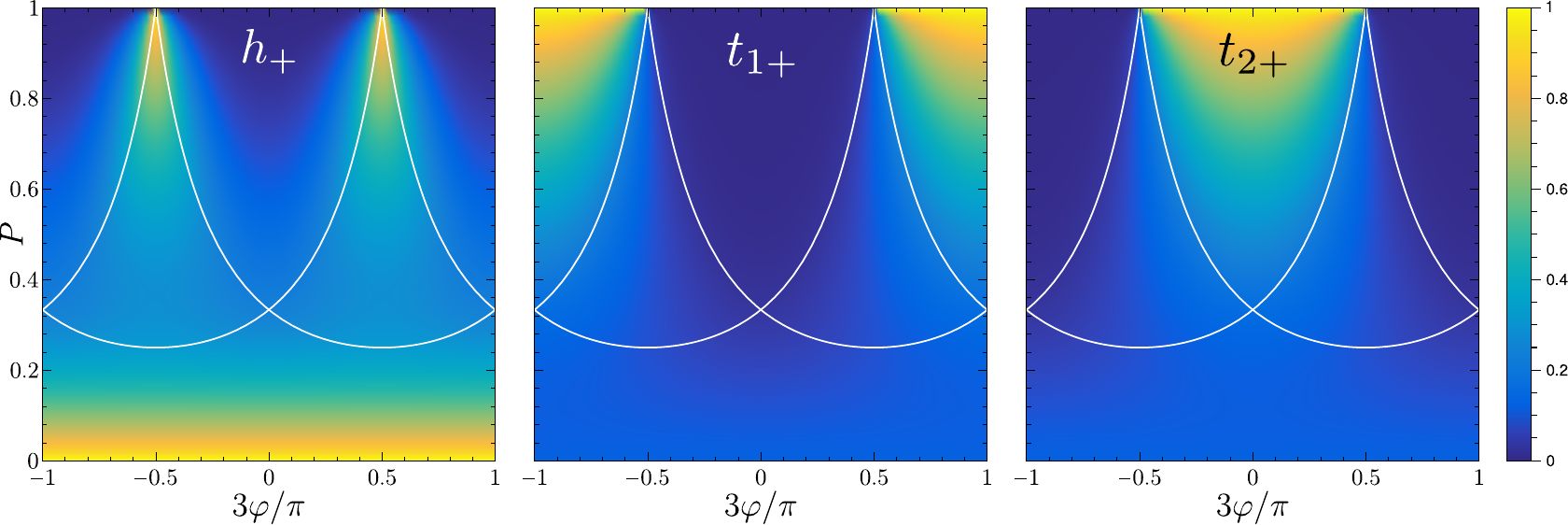}
    \caption{Loop order parameters for the band $E_{n,+}$ shown in the $(\varphi,P)$ scattering parameter space of the chiral kagome network, where the white curves give the gap closing lines.}
    \label{fig:looporder}
\end{figure}

Since there are two nonequivalent triangles in the kagome lattice, as $\mathcal C_{2z}$ is broken, there are tree possible classical loop configurations, depending on whether the loops are going along one of two nonequivalent triangles or the hexagon. On average, for a single band, the loops run only along one of the two triangles for $P \approx 1$. To this end, we calculated the following ``loop order parameters'' for the network bands $E_{n,s}(\bm k)$:
\begin{align}
    t_{1,s} & = \frac{3^3}{N} \sum_{\bm k} \left| b_{11} b_{21} b_{31} \right|^2_{s}, \label{eq:t1} \\
    t_{2,s} & = \frac{3^3}{N} \sum_{\bm k} \left| b_{12} b_{22} b_{32} \right|^2_{s}, \label{eq:t2} \\
    h_s & =  \frac{6^6}{N} \sum_{\bm k} \left| b_{11} b_{12} b_{21} b_{22} b_{31} b_{32} \right|^2_{s}, \label{eq:h}
\end{align}
which are shown in Fig.\ \ref{fig:looporder}. Here, $t_{1s}$ ($t_{2s}$) is close to unity if the band corresponds to a classical loop configuration along the downward (upward) pointing triangles in the kagome network and $h_s$ is close to unity if it corresponds to loops around hexagons (see Fig.\ \ref{fig:phase}). We find that $h_+ = h_-$ and $t_{1,s} = t_{2,-s}$. Note also that $t_{1,s}$ and $t_{2,s}$ do not vanish entirely in the $H$ phase, because in that case, the amplitude is equal for all links, and they attain the average value $3^3 / 6^3 = 1/8$.\\

\subsection{Network ribbons}

The calculation for the network ribbons goes as follows: one first constructs a large $S$ matrix that describes scattering within the unit cell of the ribbon. Next, one constructs the matrix $\mathcal M(k)$ which contains the Bloch phases. We then obtain
\begin{equation}
    \mathcal S \mathcal M(k) b_k = e^{i2\pi E/E_L} b_k,
\end{equation}
where now $\mathcal S$ contains many copies of $S_0$ together with boundary $S$ matrices for the top and bottom edge. The connectivity matrix $\mathcal M(k)$ for the ribbon is obtained in a similar way as for the bulk system; see Eqs.\ \eqref{eq:conn1}--\eqref{eq:conn2}.

\subsection{Energy dependence of the $S$ matrix}

When the $S$ matrix depends on the energy, Eq.\ \eqref{eq:network} becomes a nonlinear eigenvalue problem, which we solve self-consistently as follows. At each momentum, we first solve Eq.\ \eqref{eq:network} by replacing $\mathcal S(E)$ with $\mathcal S(E_1)$ where $E_1$ is our first guess. We then obtain energies $\{E_{2,n}\}$ where $n$ labels the eigenenergies from the lowest to the highest value. For each $n$, the energy is updated by setting $\mathcal S = \mathcal S(E_{2,n})$ and calculating the $n$th eigenvalue. Then we repeat this procedure until convergence.

The energy dependence of the scattering parameters that we have used to obtain the network band structure shown in Fig.\ \ref{fig:nbands2}(a) is explicitly given by
\begin{equation}
    P(E) = \frac{1 + f(E) - f(0)}{4 \sin^2\left( \varphi + 2\pi/3 \right)}, \qquad \varphi = -\frac{\pi}{6},
\end{equation}
with $f(E) = \exp\left(-a|3E/E_L-b|\right)$ and where we take $a = 1.2$ and $b=-0.1$.

\bibliography{references}

\end{document}